\documentclass{PHYEAUTH}
\usepackage{graphicx}
\usepackage{amsmath}
\usepackage{amssymb}

\begin{document}

\begin{frontmatter}

\title{ Magnetoresistance of Si(001) MOSFETs with high
concentration\\ of electrons}

\author[address1]{ L.\ Smr\v{c}ka \thanksref{thank1}},
\author[address1]{ O.\ N.\ Makarovsky \thanksref{thank2}},
\author[address2]{S.\ G.\  Schemenchinskii}, 
\author[address1]{P.\ Va\v{s}ek}, and
\author[address1]{V.\ Jurka}
\address[address1]{Institute of Physics ASCR, Cukrovarnick\'a 10, 162
53 Praha 6, Czech Republic}
\address[address2]{Institute for Metrological Service, 2 Andreevskaya
Nab., 117 334 Moscow, Russia}

\thanks[thank1]{
Corresponding author. 
E-mail: smrcka@fzu.cz}
\thanks[thank2]{present address: University of Nottingham, University
Park, Nottingham NG7 2RD, UK
}
\begin{abstract}
We present an experimental study of electron transport in inversion
layers of high-mobility Si(001) samples with occupied excited
subbands. The second series of oscillations, observed in addition to
the main series of Shubnikov-de Hass oscillations, is tentatively
attributed to the occupation of a subband associated with the $E_{0'}$
level. Besides, a strong negative magnetoresistance and  nonlinear
field dependence of the Hall resistance accompany the novel
oscillations at high carrier concentrations. The heating of the 2D
electron layers leads to suppression of the observed anomalies.
\end{abstract}

\end{frontmatter}
\section{Introduction}
Silicon is an indirect gap semiconductor with six minima (valleys) in
the conduction band. In the case of (001)-oriented wafers, the long
axes of two of the Fermi ellipsoids are perpendicular to the $x-y$
plane of a $\rm Si/SiO_2$ interface (with $m_z = 0.916$), while the
long axes of the other four have an in-plane orientation ($m_z =
0.19$). A quantum well is formed at the interface by applying a
positive voltage $U_g$ to the gate, and the electron motion in $z$
direction is quantized. The difference in $m_z$ causes a splitting of
the energy spectrum of bound states into two independent ladders of
levels: the twofold-degenerate ladder of eigenenergies $E_0, E_1, E_2,
\ldots$ and a fourfold-degenerate ladder $E_{0'}, E_{1'}, E_{2'},
\ldots$. Due to the higher effective mass in the $z$-direction, the
lowest energy level in the potential well is $E_0$.

The self-consistent calculations (see e.g \cite{ando}) predicts that
the excited levels $E_{0'}$ and $E_1$ are very close on energy scale
and become occupied (cross the Fermi energy $E_F$) at the carrier
concentration $N \approx 0.5 \times 10^{13} \rm cm^{-2}$ .  While
$E_F- E_1$ increases linearly with $U_g$, the fourfold-degenerate
$E_{0'}$ stays ``pinned'' to the Fermi energy and $E_F-E_{0'}$ is
almost constant even for the highest possible gate voltages. In both
cases the difference between the Fermi energy and the eigenenergies of
excited states  is at most a few percent of $E_F-E_0$, the difference
between the Fermi energy and the ground level, for all gate voltages.
\section{Samples}
In our experiments we have employed the Hall bar samples of Russian
provenance, with 200 nm thick gate oxide and the top mobilities $\mu$
above $20000 \rm cm^2/Vs$ at liquid helium temperature. The
concentration of electrons in the inversion layer is related to the
gate voltage by $dN/dU_g \approx 1.1 \times 10^{11}\rm cm^{-2}V^{-1}$
with the threshold voltage close to $ 0.5 V$. The samples are 0.25 mm
wide and 0.5 mm long with the distance between potential leads 0.625
mm.  The highest concentration of carriers $N \approx 1.3 \times
10^{13} \rm cm^{-2}$ is reached for $U_g=120 \rm V$. This is well
above  $U_g=45 \rm V$, the onset voltage of the occupation of excited
subbands. According to our numerical modeling of the electronic
structure \cite{jungw}, $E_F- E_0 \approx 70\, \rm meV$ while $E_F-
E_{0'} \approx 1.5\, \rm meV$ at the highest concentration.

The samples exhibit  the previously unreported
features, which turned out to be important in experiments with high
carrier concentration.
\begin{figure}[ht]
\begin{center}\leavevmode
\includegraphics[width=7cm]{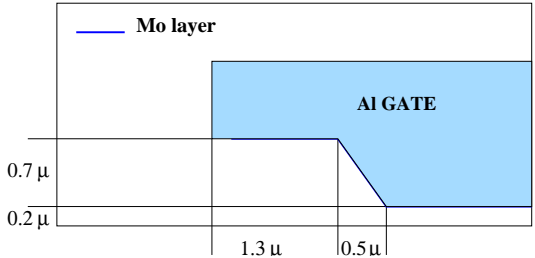}
\end{center}
\caption{The shape of the gate edge as reconstructed from the
cross-section electron microscopy. The ratio of thickness of  two
oxide layers is 9 : 2. }
\label{sample}
\end{figure}

The metallic (aluminium) gate of samples overlaps to the thicker oxide
outside the lithographically defined Hall bar shape.  The thin
molybdenium layer separates the aluminum gate and the silicon oxide.
The Mo layer can be seen by cross-sectional electron microscopy
(Fig. \ref{sample}) and allows  identification of  a step in the thickness
of oxide layers.

Therefore, a parallel channel is formed at the sample edges.  This
explains two series of magnetoresistance oscillations at high
concentration of carriers we reported in our previous publications
(see \cite{sem} and references therein).  An attempt to attribute the
second series to the SdH oscillations of electrons from the second
subband turned out to be wrong, the ratio of periods of the two series
of SdH oscillation was 9 : 2, i.e. it corresponds to the ratio of the
two oxide layers.

To suppress the effect of parallel edge-channels, approximately $10\,
\mu\rm m$ of the gate metal was lithographically removed on the
sample edges. After this procedure the second series of SdH
oscillations, corresponding to the sample edge channels, disappeared.
The high mobility of samples was not reduced by this procedure.
\section{Experiments}
\begin{figure}[th]
\begin{center}
\includegraphics[width=6.5cm,angle=-90]{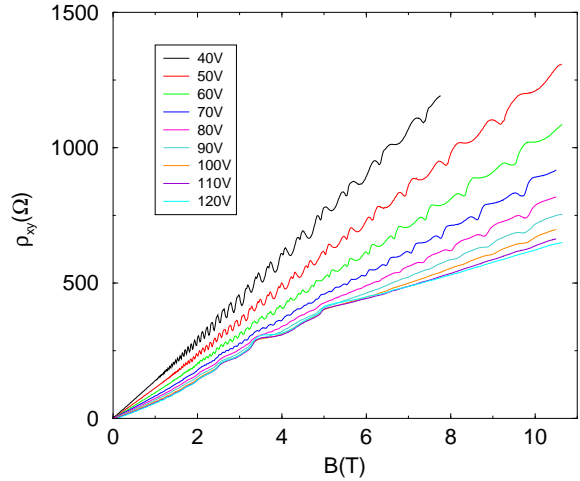}
\end{center}
\caption{The Hall resistance of Sample A for different gate voltages
deviates from the linear dependence for large $U_g$. $I_{AC} = 
0.5~\mu\rm A$, $T= 0.45~\rm K$.}
\label{hallA}
\end{figure}

\begin{figure}[h]
\begin{center}
\includegraphics[width=6.5cm,angle=-90]{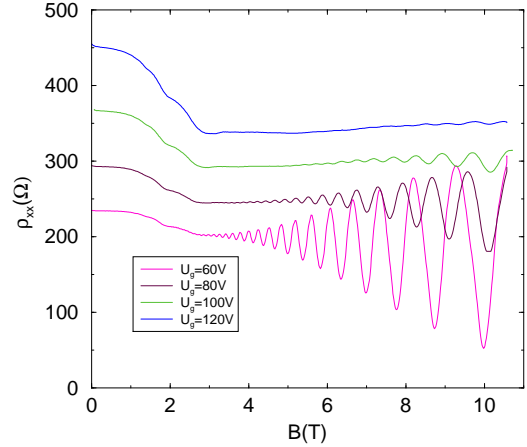}
\end{center}
\caption{The Sample B magnetoresistance of at selected gate voltages
with two series of SdH oscillations. $I_{AC} = 0.5~\mu\rm A$, $T=
0.45~\rm K$.}
\label{resB}
\end{figure}

\begin{figure}[ht]
\begin{center}
\includegraphics[width=6.5cm,angle=-90]{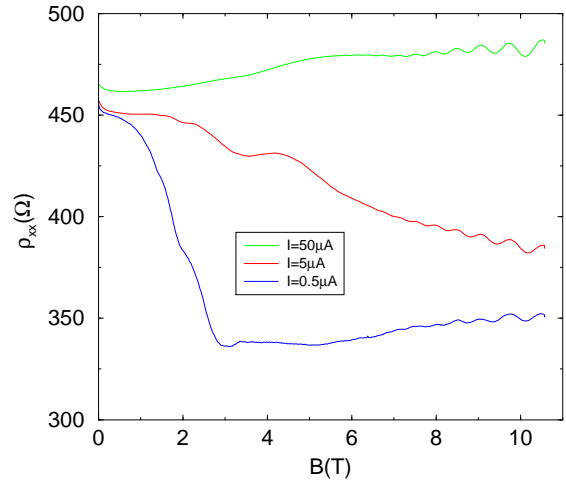}
\end{center}
\caption{The magnetoresistance of Sample B with the gate voltage
$U_g=120~\rm V$ measured at $T= 0.45~\rm K$.  Three current densities
are used to demonstrate the effect of overheating of the 2D electron
layer.}
\label{current}
\end{figure}

\begin{figure}[ht]
\begin{center}
\includegraphics[width=6cm,angle=-90]{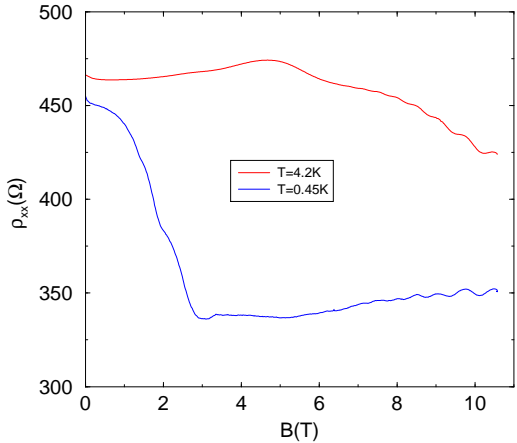}
\end{center}
\caption{The magnetoresistance of Sample B with the gate voltage
$U_g=120~\rm V$ measured at two different temperatures, $I_{ac}=
0.5~\mu$A.  The effect of the overheating of 2D electron layer by bath
is shown.}
\label{temperature}
\end{figure}

The magnetoresistance of Si(001) MOSFETs with high density of
electrons in the inversion layer was measured as a function of the
magnetic field $B$ up to 11~ T for a series of fixed gate voltages in
the bath of pumped $^3$He. The standard ac technique was used with a
frequency 13~ Hz. The measuring current $I_{AC} = 0.5~\mu\rm A$
yields the current density $\approx 20~\mu\rm A/cm$.
\begin{figure}[ht]
\begin{center}
\includegraphics[width=6.5cm,angle=-90]{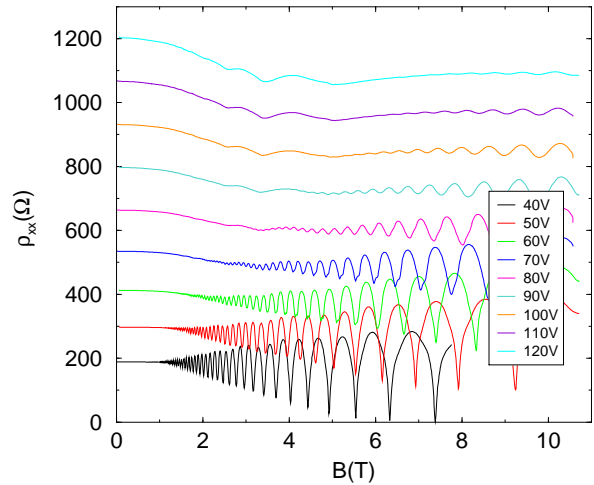}
\end{center}
\caption{The Sample A magnetoresistance  for different gate voltages
with two series of SdH oscillations. $I_{AC} = 0.5~\mu\rm A$, $T=
0.45~\rm K$. Starting from magnetoresistance at $U_g=50$~V, the
curves are offset  by multiples of $100~\Omega$ along the vertical
axis. }
\label{resA}
\end{figure}
Here we report results obtained for two samples. The Sample A has a
top mobility $\mu = 26000 \rm cm^2/Vs$ at 0.45~K. The
magnetoresistance and the Hall resistance of this sample are presented
in Figs.\ \ref{resA} and \ref{hallA} for a dense set of gate voltages.
The Sample B magnetoresistance   (the top mobility $\mu = 22000~\rm
cm^2/Vs$ is shown in Fig. \ref{resB} only for selected gate voltages
$U_g$. For this sample we also show the current-density and
temperature dependence of the magnetoresistance at the maximum gate
voltage $U_g=120~\rm V$ (see Figs. \ref{current} and \ref{temperature}).

Anomalous behavior was observed for all gate voltages corresponding to
occupied excited subbands.  

First, a novel series of SdH oscillations appeared, most pronounced at
the highest gate voltage 120~V where the amplitude of SdH oscillations
of the first subband is very small. The oscillations are periodic in
$1/B$. Their period is almost independent of $U_g$ and close
to~$\approx~0.2~ \rm T^{-1}$ for both samples.

Second, a strong negative magnetoresistance and the nonlinear field
dependence of the Hall resistance accompany the novel oscillations at
high carrier concentrations.

Both the  novel oscillations and the negative magnetoresistance are most
pronounced at the lowest temperatures ($T\approx$ 0.4~K) and the
smallest density of current ($I_{ac}= 0.5~\mu$A) used in our
measurements.

The influence of the current density on the observed effect is
illustrated in Fig.\ \ref{current}. The current was first increased by
an order of magnitude and then by two orders of magnitude. Both
amplitudes of the novel oscillations and the negative
magnetoresistance are gradually suppressed, while the amplitude of SdH
oscillations from the main series remain unchanged.

An independent measurement, in which the current density was kept low
and the bath temperature was increased to 4.2 K, confirmed that the
suppression of amplitudes of the novel oscillations and of the negative
magnetoresistance is really due to the overheating of 2D electron gas in the
inversion layer. Again, the amplitude of oscillations from the main
series are unchanged.

\section{Results and discussion}
We attribute the observed anomalies to the large difference between
the electron structure of the ground subband and excited subbands.
  
The twofold-degenerate ground subband contains many Landau levels even
for highest magnetic fields. The electron mobility in this subband
quickly decreases with increasing gate voltage. The roughness of
Si/SiO interface dominates the electron scattering. The Dingle
temperature is well above the electron layer temperature as witnessed
by the temperature independent amplitude of the SdH oscillations. The
quasiclassical picture of the electric conductivity seems to be
relevant.

The occupation of excited subbans is very small and only a small
number of Landau levels should be occupied even at relatively week
$B$.

The electrons in the subband attached to the level $E_1$ have an
effective mass 0.19, the same as the electrons in the ground subband.
With the difference $E_F -E_1$ of a few meV at high concentration
limit, the first Landau level can cross the Fermi energy at relatively
week $B \approx 3-4~\rm T$. Then the subband is emptied and the
density of states of electrons belonging to the twofold-degenerate
ladder is halved. This can contribute to reduction of the scattering
rate and to the negative magnetoresistance.

On the other hand, we believe that the fourfold-degenerate subband is
not emptied by increasing $B$ and that the level $E_{0'}$ stays pinned
to the Fermi energy also in the presence of the magnetic field. We
suggest that it is responsible for the novel oscillations the period
of which does not depend on $U_g$. The Landau level separation   in
this subband is smaller due to the larger cyclotron effective mass
$\approx$0.43. Therefore, the amplitude of oscillations should be more
sensitive to the temperature, in agreement with the experimental data.
The experiments also indicate that the Dingle temperature of these
electrons is rather low.  Assuming the fourfold degeneracy, the
concentration of electrons corresponding to $\Delta 1/B\approx 0.2~\rm
T^{-1}$ would be close to $1\times 10^{12}~\rm cm^{-2}$.

Two very different groups of electrons contribute to the
magnetoresistance of a sample with high electron concentrations: the
electrons from the ground subband and the more mobile electrons from
the excited subbands. While the transport mediated by the ground state
electrons should by described quasiclassically, the quantization of
electron states is important for excited subbands. Therefore, the deviations
from the standard quasiclassical model of transport by two groups of carries 
should be anticipated.

With increasing  temperature the energy $k_BT$ becomes comparable
with the separation between Landau levels. The electron-electron
scattering increases, the importance of a small group of electrons from
the excited subband decreases and the observed anomalies are
suppressed.

\section{Acknowledgement}
This work was supported by AVOZ 1-010-914 and in part by the Grand
Agency of the Czech Republic under Grants No. 202/01/0754 and
202/96/036.

\end{document}